\documentstyle{article}

\newcommand{\be}{\begin{equation}}
\newcommand{\ee}{\end{equation}}
\newcommand{\bea}{\begin{eqnarray}}
\newcommand{\eea}{\end{eqnarray}}
\newcommand{\nn}{\nonumber}
\newcommand{\bean}{\begin{eqnarray*}}
\newcommand{\eean}{\end{eqnarray*}}
\newcommand{\half}{\frac{1}{2}}
\newcommand{\integ}[2]{\int\limits_{#1}^{#2}\!\!}
\newcommand{\del}{\partial}
\newcommand{\dslash}{\partial\hspace{-.55em} /}
\newcommand{\dm}{\partial_{\mu}}
\newcommand{\dn}{\partial_{\nu}}
\newcommand{\ds}{\partial_{\sigma}}

\newcommand{\gm}{\gamma_{\mu}}

\newcommand{\gr}{\gamma_{\rho}}
\newcommand{\gum}{\gamma^{\mu}}

\newcommand{\e}{\varepsilon}

\newcommand{\gmnd}{g_{\mu\nu}}
\newcommand{\expect}[1]{\langle #1 \rangle}
\newcommand{\dd}{\mbox{d}}

\newcommand{\gE}{\gamma_E}
\newcommand{\ordo}[1]{{\cal O}(#1)}
\newcommand{\myref}[1]{(\ref{#1})}
\newcommand{\idk}{\int\frac{\dd^dk}{(2\pi)^d}}

\newcommand{\la}{\lambda^a}

\newcommand{\fabc}{f^{abc}}
\newcommand{\dabc}{d^{abc}}

\newcommand{\secI}[1]{\section{#1}}
\newcommand{\secII}[1]{\subsection{#1}}

\newcommand{\jma}{j_{\mu}^a}
\newcommand{\jmb}{j_{\mu}^b}
\newcommand{\jmc}{j_{\mu}^c}
\newcommand{\jna}{j_{\nu}^a}
\newcommand{\jnb}{j_{\nu}^b}
\newcommand{\jnc}{j_{\nu}^c}
\newcommand{\jra}{j_{\rho}^a}
\newcommand{\CC}{C_{\mu\nu}^{\rho}}
\newcommand{\DD}{D_{\mu\nu}^{\sigma\rho}}
\newcommand{\bPsi}{\bar{\psi}}

\begin{document}
\setlength{\baselineskip}{4.5ex}

\title{
Massive current algebra in the many-flavor chiral Gross--Neveu model}
\author{
  \normalsize{ Tamas Hauer\thanks{ E-mail: hauer@ludens.elte.hu }}
  \\ \\
  \normalsize{ Institute for Theoretical Physics}\\
  \normalsize{ Lorand E\"otv\"os   University} \\
  \normalsize{ H-1088 Budapest, Puskin u. 5-7, Hungary}
}
\def\today{}
\maketitle

\begin{abstract}
We study the algebra of $SU(n)$-currents in the many-flavor chiral
Gross-Neveu model. The general structure of the current-current OPE leading
to non-local quantum conserved charges is reviewed. We calculate the
OPE in the one-flavor and the many-flavor models perturbatively and use
renormalization group invariance to prove that our results are not altered
by higher-order corrections. We conclude that in these models the non-local
quantum charge exists which is the first step towards the proof of the
absence of particle production and factorization.
\end{abstract}

 
keywords: non-local charge, operator product expansion, Gross-Neveu model, 
          renormalization group

{\flushleft Address of correspondence: \\
Tamas Hauer \\
Institute for Theoretical Physics, Lorand E\"otv\"os   University \\
H-1088 Budapest, Puskin u. 5-7, Hungary \\
Tel: 36-1-266-7921; Fax: 36-1-266-0612 \\
email: hauer@ludens.elte.hu}

\newpage

\secI{Introduction}

Two dimensional integrable systems  -- being
exactly solvable -- are ideal toy models for understanding the behavior of
higher dimensional
quantum field theories.  In many cases a rich symmetry structure is hidden behind
integrability and the existence of a sufficient number of conserved charges
provides a non-perturbative definition of the model. These conservation laws can
yield the determination of the exact S-matrix and form factors. A large class
of massive integrable QFT's are characterized by infinite
dimensional non-abelian quantum group symmetries (for a review see
\cite{BerLecl}). The elements of the underlying Yangian algebra are
conserved charges which may be obtained as spectral coefficients of the
quantum monodromy operator \cite{devega1,devega2}, alternatively the generators can be 
defined as non-local functions of the local symmetry currents.
The non-perturbative definition of such a non-local charge was first
constructed by L\"uscher in the non-linear $\sigma$-model \cite{Luscher}.  Later
other models were also investigated \cite{Bernard,Abdalla,AbAb}.  In all cases
the short-distance expansion of the product of the local currents provides the
key to the quantum conservation laws:  different model-specific assumptions finally
lead to the theorem proved in L\"uscher's original paper.  Possible
generalizations to higher dimensions is discussed in \cite{Buchholz,Buchholz2}.

As it was pointed out by Bernard \cite{Bernard} the origin of the Yangian
symmetry in the quantum theory is the quantum version of the curl-free equation
of the local currents:
\bea
\label{classcurlfree}
\del_0j_1^a-\del_1j_0^a+\fabc j_0^bj_1^c=0.
\eea
If this field equation holds then the following expression defines a conserved
quantity in the classical theory:
\bea
\label{classcharge}
Q_1^a = \frac{1}{4}\integ{-\infty}{\infty} dy_1dy_2\epsilon(y_1-y_2)\fabc
j_0^b(t,y_1)j_0^c(t,y_2) +
\integ{-\infty}{\infty}dy j_1^a(t,y)\,.
\eea

In the quantum theory both of these equations are modified.  One has to
introduce an appropriate normal ordering to define the composite operator
appearing in the curl-free equation which may lead to quantum corrections of
this relation. The quantum non-local charge has to be defined using some
regularization.  The natural scheme is point-splitting which calls our
attention to the short-distance behavior of the product of local operators.
Indeed it turns out that it is the operator product expansion (OPE) of the
local currents which should be viewed as the quantum analog of
\myref{classcurlfree}; it provides a natural normal-ordering and also the proper
definition of the singular non-local charge.

Once we have been able to define the conserved $Q_1^a$'s, the whole Yangian
algebra can be generated by commutators of $Q^a$'s and $Q_1^a$'s
\cite{BerLecl}.  The highly restrictive nature of this rich
symmetry has been studied both in general and in specific models (see
e.g.  \cite{Bernard2,LeclSmir}).  
Since the non-local charges do not commute with the
Lorentz-boost operator, the Poincar\'e-algebra is no longer a direct multiplier
of the symmetry.  This allows one to deduce relations among the masses of the
spectrum of particles
(from the nontrivial comultiplication applied to multiparticle
scattering) \cite{Belavin}.  Also one can prove the absence of particle
production and factorization which leads to the exact S-matrix \cite{Luscher}.

In this paper we study the above OPE.
Our aim is to formulate a general criterion
for the existence and conservation of the L\"uscher-type
non-local charge. Starting from the general
OPE within the framework of asymptotically free theories, we find that the
quantum counterpart of the non-local expression \myref{classcharge} exists and it
is conserved if and only if the OPE contains no operators besides the
current and its derivative. This is essentially the summary of the results of
previous investigations.
We first study how general principles constrain  -- besides the
coefficient of the current and its derivatives -- the presence and
coefficients of possible extra operators.
The general result is the following: apart from the current and its
derivative, antisymmetric Lorentz-tensor operators may appear at the
right hand side of the OPE with constant coefficients.
This statement is important, because -- as we
will see in the concrete example -- this makes it possible to decide whether a
given operator is present or not.

As an application of the general results we present our calculation in a
special case, the multiflavor chiral Gross-Neveu model.
Here we are able to derive exact OPE coefficients  thus prove the
conservation of the quantum non-local charge which generates the Yangian
algebra. This proves that the multiflavor CGN is a concrete realization of
Bernard's massive current algebra \cite{Bernard}.

The plan of the paper is as follows.  In section 2 we study the
general features of the operator product $\fabc\jmb(x)\jnc(0)$.
We determine the most general form of the expansion.  In section 3
we first show that the chiral Gross-Neveu model satisfies the
requirements we set at the beginning of the general analysis then apply these
results to it.  We calculate the OPE perturbatively up to the first nontrivial
order and argue that it is not modified at any higher order in perturbation
theory.

\secI{The operator product expansion }
\label{gensec}

The first part of this paper is devoted to the study of the Wilson-expansion of
the product of local symmetry currents.  After fixing the family of the 2d
models we plan to work in (which is intended to be general enough to include
the formerly investigated theories) we deal with three questions:  the operator
content of the algebra, the functional form of the coefficients and the
relation of these to the existence of the non-local charge. Examination of
these points in detail yields, as a conclusion, the most general form of
the expansion in the given family.  We will show that concrete field theories
are characterized by one scalar function and some constant numbers, and that
the vanishing of the latter ensures the conservation of the non-local charge.

\secII{The class of models}

The basis of the hierarchy of the non-local charges is the algebra of local
currents, therefore we require a model to contain a set of local conserved
currents, $j^{a\mu}(x)$, with the corresponding charges forming the following
algebra:
\bea Q^a & \equiv & \int dx j^{a0}(x,t) \nn \\
\lbrack Q^a,Q^b \rbrack &=& \fabc Q^c. \label{algebra}
\eea
Here $f^{abc}$ are structure constants of a semisimple Lie-algebra, {\cal
g}\footnote{Conventions:  $f^{abc*} = -f^{abc} \;\;\; f^{abc}f^{bcd} =
-2q\delta^{ad}.$}.  We also require that the QFT is renormalizable (to give
meaning
to the OPE) and asymptotically free (to carry out dimensional analysis).  Our
main additional assumption is the following:  {\em In the adjoint
representation of} {\cal g} {\em the current $\jma(x)$ is the only operator of
dimension$<$2.} When composing this condition we kept the well known specific
models in view and we think that their similar behavior can be traced back to
this property.  Examples are the $O(n)$ non-linear $\sigma$-models -- where
because of the constraint on the length of the isovector field new composite
operators can be constructed by application of derivatives only -- and
Lagrangian fermionic theories -- where the basic field is dimensionful.
Another approach is the framework of the massive current algebras
\cite{Bernard} where the smooth UV limit directly constrains the operator
content.

From now on we will assume that these properties are valid in which case we
can write down the most general form of our Wilson-expansion:
\bea
\fabc \jmb(x)\jnc(0) &=& C^{\rho}_{\mu\nu}\jra(0) +
                 D^{\sigma\rho}_{\mu\nu}(x)\ds\jra(0) + \ldots
\label{general} \\
C_{\mu\nu}^{\rho}(x) &=& \ordo{|x|^{-1-0}} \nn\\
D_{\mu\nu}^{\sigma\rho}(x) &=& \ordo{|x|^{-0}}, \nn
\eea
where $\ldots$ include other terms of $\ordo{|x|^{-0}}$ and we denote with
$-0$ the possible logarithmic (i.e. weaker than any power) corrections to the
power-like singular behavior. In the following we will study the set of
operators on the rhs and the general form of their coefficients by
(following the idea of \cite{Luscher}) making use of general principles:
discrete symmetries, locality, Lorentz-covariance and current conservation.

\secII{Operator content: discrete quantum numbers}

Let's denote charge conjugation (on the Hilbert space) by $C$, parity by
$P$,
and CPT by $\Theta$.  The transformation properties of the current can be
summarized as:
\bea
\Theta \jma(x,t)\Theta^{-1} &=& -\jma(-x,-t)
\label{CPTcur} \\
\Theta \dm\jna(x,t)\Theta^{-1} &=& \dm\jna(-x,-t) \nn \\
P \jma(x,t) P^{-1} &=& j^a_{\tilde{\mu}}(-x,t) \nn \\
P \dm\jna(x,t) P^{-1} &=& \partial_{\tilde{\mu}}j^a_{\tilde{\nu}}(-x,t) \nn \\
C \jma(x,t) C^{-1} &=& \beta^{ab}\jmb(x,t).
\label{chargec}
\eea
Here tilde above a Lorentz index means multiplication by 1 or -1 if it is
0 or 1, respectively. The algebra of the charges requires the following
identities to hold for the matrix representing the charge conjugation:
\bean
\beta^{ab}\beta^{bc} &=& \delta^{ac} \\
\beta^{ai}\beta^{bj}f^{ijc} &=& f^{abk}\beta^{kc}.
\eean
Now consider the general expression for the OPE:
\bea
\fabc\jmb(x)\jnc(0) = \sum C_{\mu\nu}^{(i)}(x)O_{(i)}^a(0).
\eea
Here $O_{(i)}$ are linearly independent, hermitian operators.
Application of the discrete symmetry transformations yields:
\bea
C O_{(i)}^a(x)C^{-1} &=& \beta^{ab}O_{(i)}^b(x) \\
C_{\mu\nu}^{(i)}(x,t)\Theta O_{(i)}^a(0)\Theta^{-1} &=&
C_{\mu\nu}^{(i)}(-x,-t)O_{(i)}^a(0)
\label{CPT} \\
C_{\mu\nu}^i(x,t)P O_i^a(0)P^{-1} &=&
C_{\tilde{\mu}\tilde{\nu}}^i(-x,t)O_i^a(0).
\eea
The first equation shows that only operators with the same charge conjugation
transformation character (identical to that of the current) may be present.
The CPT and P symmetry involve the transformation of $x$, too, thus we get
constraints on the coefficients in which the CPT-, and the P-charge of the
corresponding operator appears.

\secII{The x-dependence of the OPE coefficients}

In this section we will analyze the possible x-dependent behavior of the
OPE-coefficients.  First we note that antihermicity of the lhs requires that
the OPE coefficients of self-adjoint operators be pure imaginary.
The analytic properties of these functions are determined by the spectrum
condition which states that they are boundary values of complex functions
analytic in $z=x-iy$ in the region $y^2>0, y^0>0$ \cite{Streater}.

As we will see in the next paragraphs the different operators do not influence
each other's behavior under the general principles. The only exception is
locality which mixes an operator with its derivative. However there is only one
operator the derivative of which may appear due to our basic assumption.
Therefore we start with examining the coefficients of the current and its
derivative first and then we turn to the other operators.

\paragraph{CPT-symmetry}

From the equations \myref{CPTcur} and \myref{CPT} we get:
\bea
C_{\mu\nu}^{\rho}(-x) &=& -C_{\mu\nu}^{\rho}(x) \\
D_{\mu\nu}^{\sigma\rho}(-x) &=& D_{\mu\nu}^{\sigma\rho}(x).
\eea

\paragraph{Locality}

For spacelike $x$ the two operators commute:  $\fabc[\jmb(x),\jnc(0)] = 0$
The
operator product expansion of this equality reads:
\bea
0 &=& \fabc\jmb(x)\jnc(0) + \fabc\jnb(0)\jmc(x) = \nn\\
&=& \left(C_{\mu\nu}^{\rho}(x) + C_{\nu\mu}^{\rho}(-x)\right)\jra(0) +
\nn\\
&&\left(D_{\mu\nu}^{\sigma\rho}(x) +D_{\nu\mu}^{\sigma\rho}(-x)
  +x^{\sigma}C_{\nu\mu}^{\rho}(-x)\right)\ds\jra(0) + O(|x|^{1-0}).
\label{locality}
\eea

This immediately yields:
\bea
C_{\nu\mu}^{\rho}(x) = C_{\mu\nu}^{\rho}(x).
\eea

To find the similar equation for $\DD$ we divide the $\sigma\rho$-tensors
into
antisymmetric, symmetric traceless and trace parts:
\bea
D_{\mu\nu}^{\sigma\rho}(x) - \half x^{\sigma}C_{\nu\mu}^{\rho}(-x)  &=&
\bar{D}_{\mu\nu}^{\sigma\rho}(x) + \tilde{D}_{\mu\nu}^{\sigma\rho}(x) +
\half g^{\sigma\rho}D_{\mu\nu}(x)   \\
\bar{D}_{\mu\nu}^{\sigma\rho}(x) &\equiv&
D_{\mu\nu}^{[\sigma\rho]}(x) - \half x^{[\sigma}C_{\mu\nu}^{\rho]}(x) \nn\\
\tilde{D}_{\mu\nu}^{\sigma\rho}(x) &\equiv&
D_{\mu\nu}^{\{\sigma\rho\}}(x) - \half x^{\{\sigma}C_{\mu\nu}^{\rho\}}(x)
-\half g^{\sigma\rho}D_{\mu\nu}(x) \nn\\
D_{\mu\nu}(x) &\equiv&
g_{\sigma\rho}\left(
D_{\mu\nu}^{\sigma\rho}(x) - \half x^{\sigma}C_{\mu\nu}^{\rho}(x)  \nn
\right).
\eea

Current conservation implies that the trace part drops out from
\myref{locality} and we obtain:
\bea
\bar{D}_{\nu\mu}^{\sigma\rho}(x) + \tilde{D}_{\nu\mu}^{\sigma\rho}(x)
=-(\bar{D}_{\mu\nu}^{\sigma\rho}(x) + \tilde{D}_{\mu\nu}^{\sigma\rho}(x)).
\eea

If there are other constraints on the tensor structure of the derivative of
the
current then this equation can be modified.  For example in case of a free
massless fermionic theory the conservation of the axial current implies that
$\ds\jra(x)$ is symmetric, in which case $\tilde{D}_{\nu\mu}^{\sigma\rho}(x)
=-\tilde{D}_{\mu\nu}^{\sigma\rho}(x)$.

\paragraph{Lorentz-covariance}

Now we apply the previously found conditions to determine the most general
Lorentz-tensor structure.  One can construct tensors from $x^{\mu}$ and
$g_{\mu\nu}$:
\bea
C_{\mu\nu}^{\rho}(x) &=& C_1g_{\mu\nu}x^2x^{\rho}+
                         C_2x^2(x_{\mu}\delta_{\nu}^{\rho}+
                                      x_{\nu}\delta_{\mu}^{\rho})+
                         C_3x_{\mu}x_{\nu}x^{\rho} \\
\tilde{D}_{\mu\nu}^{\sigma\rho}(x) &=& D_S\left(
   x^{\sigma}(x_{\mu}\delta_{\nu}^{\rho}-x_{\nu}\delta_{\mu}^{\rho}) +
x^{\rho}(x_{\mu}\delta_{\nu}^{\sigma}-x_{\nu}\delta_{\mu}^{\sigma})\right)\\
\bar{D}_{\mu\nu}^{\sigma\rho}(x) &=&
   D_A(\delta_{\mu}^{\sigma}\delta_{\nu}^{\rho}-
                \delta_{\nu}^{\sigma}\delta_{\mu}^{\rho})x^2,
\eea

where $C_1, C_2, C_3, D_S, D_A$ are scalar functions depending on $-x^2$. The
$C$'s are of order $|x|^{-4-0}$ while $D_S$ and $D_A$ both are of
$|x|^{-2-0}$.

\paragraph{Current conservation}

We write down differential equations for the scalar
functions defined above which are implied by current conservation. It is clear
that in order the current be conserved the following equations have to
hold:
\bea
\del^{\mu}C_{\mu\nu}^{\rho} = \del^{\mu}D_{\mu\nu}^{\sigma\rho} = 0.
\eea

It is straightforward to rewrite this equation in terms of the scalar
coefficients. We get more elegant forms if we introduce
new scalar functions which possess only logarithmic singularities:
\bean
C_i(-x^2) &\equiv& \frac{\gamma_i(\log(-\mu^2x^2))}{(-x^2)^2} \\
D_S(-x^2) &\equiv& -\frac{1}{4}\frac{\delta_S(\log(-\mu^2x^2))}{(-x^2)} \\
D_A(-x^2) &\equiv& -\frac{1}{4}\frac{\delta_A(\log(-\mu^2x^2))}{(-x^2)},
\eean
where we introduced an arbitrary mass scale, $\mu$.
The differential equations for these functions are:
\bean
\dot{\gamma}_1+\dot{\gamma}_2+\dot{\gamma}_3 - (\gamma_1+\gamma_2) &=& 0 \\
2\dot{\gamma}_2 + (\gamma_1+\gamma_2) &=& 0 \\
2\dot{\delta}_S + (\gamma_1+\gamma_2)  &=& 0 \\
2\delta_S + (\gamma_1+\gamma_2+\gamma_3) &=& 0 \\
2\dot{\delta}_A - \gamma_1 +\gamma_2 &=& 0,
\eean
where dot above a function means differentiation with respect to the
variable $\log(-\mu^2x^2)$.

The set of equations can be solved up to an undetermined function. First
note that the combination
\bea
\label{lambdasol}
2\lambda \equiv \gamma_1+3\gamma_2+\gamma_3.
\eea
is constant. We can express $\gamma$'s in terms of $\delta_S$:
\bea
\gamma_1 &=& -2\dot{\delta}_S-\delta_S-\lambda \nn\\
\gamma_2 &=& \delta_S + \lambda \\
\gamma_3 &=& 2\dot{\delta}_S -2\delta_S, \nn
\eea
and we can write all the five coefficients in terms of one undetermined
function, $\xi(\log(-\mu^2x^2))$:
\bea
\delta_S &=& -\dot{\xi}-\lambda \\
\delta_A &=& \dot{\xi}+\xi+\lambda.
\eea
Note that all the model dependent information is encoded in $\xi$.

\secII{Commutators of the conserved charges}

Before studying the OPE any further (i.e. the presence of operators not
considered so far) we now verify that the operator product we have obtained
this way
is consistent with the algebra of the charges, defined in eq \myref{algebra}.
We also want to relate the constant $\lambda$ to the normalization of the
algebra. One can do this by calculating the (antisymmetrized) equal-time
commutators of the currents, $\jma$ in terms of the invariant functions for
small x.
\bea
\fabc [\jmb(x),\jnc(0)]_{ET} = \lim_{\e\rightarrow 0} \fabc \left(
\jmb(x,-i\e)\jnc(0)-\jmb(x,i\e)\jnc(0)\right).
\eea
We substitute the scalar OPE-coefficients at argument $(-\e^2-x^2)$.
The terms proportional to $\dm\jna$ are zero as $\e\rightarrow 0$ and the
commutators become:
\bea
\fabc [j_0^b(x),j_0^c(0)]_{ET} &=& \lim_{\e\rightarrow 0} \left\{
2i\e\left((C_1+2C_2+C_3)\e^2+(C_1+2C_2)x^2\right)j_0^a(0)\right\} \nn\\
\fabc [j_0^b(x),j_1^c(0)]_{ET} &=& \lim_{\e\rightarrow 0} \left\{
2i\e\left(C_2\e^2 + (C_2+C_3)x^2\right)j_1^a(0)\right\}
\label{commrel}\\
\fabc [j_1^b(x),j_1^c(0)]_{ET} &=& \lim_{\e\rightarrow 0} \left\{
(-2i\e)\left(C_1\e^2 + (C_1+C_3)x^2\right)j_0^a(0)\right\}. \nn
\eea
We require the currents to be normalized according to \myref{algebra}:
\bea
\fabc[j_0^b(x),\jmc(0)]_{ET} = -2q\delta(x)\jma(0).
\eea
The normalization of the $C_i$'s can be easily extracted by noting that the
first and second commutator together yields:
\bea
-4q\delta(x) = \lim_{\e\rightarrow 0}
\{ (\gamma_1+3\gamma_2+\gamma_3) \frac{2i\e}{\e^2+x^2} \},
\eea
which gives:
\bea
\lambda &=& -\frac{q}{i\pi}.
\eea

Let's examine the commutator of the spacelike components of the current,
too.
Eq. \myref{commrel} shows, that it is proportional to the timelike component.
We introduce the (model dependent) constant $\omega$ with:
\bea
\fabc[j_1^b(x),j_1^c(0)]_{ET} = -2q\omega\delta(x)j_0^a(0).
\label{spacelike}
\eea
The two commutators involving $j_1^a$ yield:
\bea
-2q(1+\omega)\delta(x) &=& \frac{2i\e}{\e^2+x^2}(\gamma_2-\gamma_1) \\
\omega&=&\frac{-2}{\lambda}\dot{\delta}_A - 1.
\label{omega1}
\eea
The scalar function $\delta_A$ is a possibly singular function of
$\log(-\mu^2x^2)$. We expand it around $x=0$ as:
\bea
\delta_A = d(\log(-\mu^2x^2))^p+\ordo{(\log(-\mu^2x^2))^{p-1+0}}.
\label{pdef}
\eea
From \myref{omega1} we see that different models fall
into three categories.  If
$p>1$ then $\omega$ is infinite, the commutator of the spacelike components is
not consistent with \myref{spacelike}.  If $p<1$ then $\dot{\delta_A}$ vanishes
as $x\rightarrow 0$ and $\omega=-1$ while if $p=1$ then
\bea
\omega = -2\frac{d}{\lambda}-1.
\label{omega}
\eea

\secII{L\"uscher's non-local charge}

We are now in the position to define the quantum counterpart of
\myref{classcharge}. The singularity is regularized by introducing a
"point-splitting charge"\cite{Luscher}:
\bea
\label{quantumcharge}
Q^a_\zeta &=&
\integ{|y_1-y_2|\geq\zeta}{} dy_1dy_2\epsilon(y_1-y_2)\fabc
j_0^b(t,y_1)j_0^c(t,y_2) + \nn \\
&&+ Z(\zeta)\integ{-\infty}{\infty}dy j_1^a(t,y).
\eea
The (divergent) renormalization $Z(\zeta)$ should be chosen so that the above
expression gives a well-defined operator in the limit $\zeta\rightarrow 0$.
It is clear that the divergence to be compensated is closely related to the
singularity of the current-current OPE. Moreover it turns out
that the existence of the limit is not influenced by the presence of extra
operators in the expansion and the proper choice for $Z(\zeta)$
involves the coefficients of the derivative of the current only:
\bea
\label{renconst}
Z(\zeta)&=&\delta_S(\log(\mu^2\zeta^2))-\delta_A(\log(\mu^2\zeta^2)).
\eea
There remains, of course the question whether the operator defined this way is
conserved. This can be answered by taking the time-derivative of the
point-splitting charge and examine if it vanishes or not in the
$\zeta\rightarrow 0$ limit. Again we have to insert the OPE in the
integral after which we find that if the renormalization constant
$Z(\zeta)$ is defined as in \myref{renconst} then
the contribution coming from those part
of the OPE which involves the current and its derivative, vanishes as
$\zeta\rightarrow 0$. On the other hand any operator which is independent
of the
current gives a nonvanishing result in this limit. This leads to the conclusion
that {\em the conservation of the non-local charge in the quantum theory
depends on the existence of extra operators in the current-current OPE only.}

\secII{Coefficients of other operators}

After investigating the OPE coefficient of the current and its derivative
the previous section provides enough motivation to turn to the question of what
other operators can appear on the rhs of the expansion.

We showed that only those operators may be present which have the
same $C$-parity as the current and due to our assumption that there is no
operator with lower dimension than that of the derivative of the current,
their coefficients are of order $O(|x|^{-0})$.

Lorentz-scalar operators are ruled out since their coefficient should be
proportional to $\e^{\mu\nu}$ - the only tensor which possesses antisymmetry,
but parity is violated then. Pseudoscalar operators can occur, their
coefficient must be proportional to $\e^{\mu\nu}$. For the coefficient,
$G_{\mu\nu}^{\rho}(x)$ of a vector operator CPT and locality yield the same
constraint as for that of the current but now it should be of dimension
$O(|x|^{-0})$. It is easy to see that there is no Lorentz-tensor of
this kind composed from $g_{\mu\nu}$ and $x^\mu$ which satisfies the
spectrum condition.
The next simplest operators are
the symmetric and antisymmetric tensors. The coefficient of a symmetric
tensor
can be described by one scalar function:
\bea
B_{\mu\nu}^{\sigma\rho}(x) = B(x^2)
x^{\{\sigma}x_{[\mu}\delta^{\rho\}}_{\nu]}.
\eea
One can easily see that this satisfies the CPT condition and locality.
However current conservation yields:
\bea
\del^\mu B_{\mu\nu}^{\sigma\rho} =
(x^\sigma\delta_\nu^\rho+x^\rho\delta_\nu^\sigma)(2x^2B'+2B) -
4x_\nu x^\sigma x^\rho B' - 2x_\nu g^{\sigma\rho}.
\eea
The only solution to this equation is the constant zero, the
symmetric tensors are also ruled out. Let us now consider
an antisymmetric tensor, the coefficient of which can again be described
by a scalar function:
\bea
E_{\mu\nu}^{\sigma\rho}(x) = E(x^2)
(\delta_\mu^\sigma\delta_\nu^\rho-\delta_\nu^\sigma\delta_\mu^\rho).
\eea
This satisfies the locality condition and CPT. The current conservation
yields the following constraint on the functional form of $E$:
\bea
\dm E = 0.
\eea
That is, any antisymmetric operator in our OPE must
have {\em constant} coefficient. Note that the same holds for a pseudoscalar
operator, since its coefficient must be proportional to $\e^{\mu\nu}$ and
after redefining the operator so that it includes $\e^{\mu\nu}$ it becomes
an antisymmetric tensor. Therefore the most general form of $\ldots$ in
\myref{general} is a series of antisymmetric tensoroperators with constant
coefficients.

\secII{Summary}

In this section we have gathered all the information about the current-current
OPE which are valid model-independently. The individual theories are
characterized by the function $\xi$ and the constants $E_i$ belonging to
operators $O_i$ as follows:
\bea
\fabc\jmb(x)\jnc(0) = \hfill \nn
\eea
\bea
\left(\gamma_1\frac{g_{\mu\nu}x^{\rho}}{x^2} +
  \gamma_2\frac{x_{\mu}\delta_{\nu}^{\rho}+x_{\nu}\delta_{\mu}^{\rho}}{x^2}
+
  \gamma_3\frac{x_\mu x_\nu x^\rho}{x^4}\right)
\left(\jra(0)+\frac{x^\sigma}{2}\ds\jra(0)\right) +
\nn
\eea
\bea
\left(\frac{\delta_S}{4}\frac{x_\mu x^\rho\delta_\nu^\sigma+
x_\mu x^\sigma\delta_\nu^\rho-x_\nu x^\rho\delta_\mu^\sigma-
x_\nu x^\sigma\delta_\mu^\rho}{x^2}
+\frac{\delta_{A}}{4}
(\delta_\mu^\sigma\delta_\nu^\rho - \delta_\nu^\sigma\delta_\mu^\rho)\right)
\ds\jra(0)
\nn
\eea
\bea
+\sum_{i}E_i O_{[\mu\nu]}^i(0),
\eea

with
\bea
\gamma_1 &=& -2\dot{\delta}_S-\delta_S +\frac{q}{i\pi} \\
\gamma_2 &=& \delta_S-\frac{q}{i\pi} \\
\gamma_3 &=& 2\dot{\delta}_S-2\delta_S \\
\delta_S &=& -\dot{\xi}+\frac{q}{i\pi} \\
\delta_A &=& \dot{\xi} +\xi -\frac{q}{i\pi} \\
E_i &=& constant.
\eea

The interesting models are in which no operators other than the current
and its derivative appear in the OPE. In these cases a non-local conserved
quantum charge can be defined as in \myref{quantumcharge},\myref{renconst}.

As the first example for the above we recall the $O(n)$ non-linear
$\sigma$-model. The leading behavior of the OPE coefficients is the
following \cite{Luscher}:
\bean
\gamma_1 = -\frac{\lambda}{2} \;\;\;
\gamma_2 =  \frac{\lambda}{2} \;\;\;
\gamma_3 = \lambda
\eean
\bean
\delta_S  &=& -\frac{\lambda}{2} \\
\delta_A &=& -\frac{\lambda}{2}(1+\log(-\mu^2x^2)) \\
\omega   &=& 0.
\eean
Thus in the $O(n)$-models $\xi=-\lambda(1+\half\log(-\mu^2x^2))$. (As it was
pointed out by L\"uscher \cite{Lerr} there must be
subleading ($\log\log()$) singularities too.) The parameter $\mu$ is
proportional to to the mass of the particles forming the $n$-plet. From the
singularity of $\delta_A$, $p=1$ (see \myref{pdef}) and substituting
it into \myref{omega} we get $\omega=0$. This is exactly the expected behavior:
the spacelike components of the current are constructed out of the spacelike
derivatives of the basic scalar field and these necessarily commute.

Now let us see how the construction based on the massive current algebra
\cite{Bernard} fits in this picture. Here the assumption is that the smooth
UV-limit of the currents leads to the corresponding Kac-Moody
algebra i.e. clearly no operator can appear in the OPE. Following
\cite{Bernard} the OPE-coefficients of the models belonging to this
category:
\bean
\gamma_2 &=& -\gamma_1 = \lambda \;\;\; \gamma_3 = 0 \\
\delta_S &=& 0 \\
\delta_A &=& -\lambda\log(-\mu^2x^2) \\
\omega &=& 1.
\eean
Here again only the leading singularity is shown. One can see that
$\xi=-\lambda\log(-\mu^2x^2)$, ($\mu$ is a model-dependent mass parameter),
again $p=1$, which now leads to $\omega=1$, that is
\bea
\fabc[j_1^b(x),j_1^c(0)]_{ET} = -2q\delta(x)j_0^a(0).
\eea
As a concrete example for these algebras Bernard studies the
$SO(n)$ GN-models -- assuming that these really belong to this family. In
the remaining part we deal with the $SU(n)$ multiflavor chiral
GN-models and prove that they are concrete realizations of the massive current
algebra framework.


\secI{The many-flavor chiral Gross-Neveu model}

In the second part of this paper we will give an example for the above results
in a concrete Lagrangian quantum field theory. Our choice is the
(one- and) many-flavor chiral Gross-Neveu model (CGN) \cite{GN}, these theories 
possess the set of conserved local currents with the algebra \myref{algebra} 
so one may expect to be able to define a non-local conserved charge. This is 
certainly the case for the one-flavor model where both the 
Ward identities corresponding to the non-local symmetry \cite{Zachos}
and the explicit construction of the quantum charge \cite{AbAb} have been studied.
As we saw in the previous section the sufficient condition for the existence
of the quantum charge is that only the currents
and their derivatives appear in our OPE. This is what we will prove in the
following. Our strategy is:  we allow all the possible operators
on the rhs of the OPE which can be constructed from the basic fermionic field
with the right quantum numbers and dimension.  Then we show that either
the corresponding coefficients are zero or the operator is not linearly
independent of the current.


\secII{The Lagrangian theory of the CGN model}

The chiral Gross-Neveu (CGN) model \cite{GN} is one of the most investigated
toy-models.  It is asymptotically free\cite{Mitter}, the masses of the particle
spectrum are generated dynamically\cite{GN}.  The continuous chiral symmetry is
unbroken:  the massive particles do not carry chiral charge\cite{Witten}.  The
extension of the model to the multiflavor case is motivated by the equivalence
of the $N_{flavors}\rightarrow \infty$ limit and the principal $SU(n)$ chiral
model\cite{PolyWieg,Destri}.  Two-loop renormalizability of the multiflavor
model is proven in \cite{Destri2}.

Though there are indications that there is much
difference between the one- and the multiflavor models they can be treated
similarly from the perturbative point of view.  Therefore in the following
section we will speak about the many-flavor model without restricting ourselves
to the number of flavors greater than one case.  The basic field is a massless
fermion with a color and a flavor degree of freedom:
\bea
\psi(x) \equiv \psi^{Ii}(x),
\nn
\eea
where $I$ runs from 1 to $N$ and transforms under the fundamental
representation of $SU(N)_f$ and $i$ runs from 1 to $n$ and transforms
under the fundamental representation of $SU(n)_c$. There are four
types of vector currents which can be built as bilinears of the basic
field. These are:
\bean
j^\mu(x)      &\equiv& \bar\psi(x)\gamma^\mu\psi(x) \\
\jma(x)       &\equiv& \bar\psi(x)\lambda^a\gamma_\mu\psi(x) \\
j_\mu^A(x)    &\equiv& \bar\psi(x)\Lambda^A\gamma_\mu\psi(x) \\
j_\mu^{Aa}(x) &\equiv& \bar\psi(x)\lambda^a\Lambda^A\gamma^\mu\psi(x).
\eean

Here $\lambda^a$ and $\Lambda^A$ are the generators of the color and
flavor group in the fundamental representation \footnote{conventions:
$[\lambda^a,\lambda^b] = \fabc\lambda^c\;\;\;f^{abc*} = -f^{abc} \;\;\;
f^{abc}f^{bcd} = -2q\delta^{ad}$.}. The Minkowskian action is:
\bea
S = \int d^2x \left(\bPsi i\dslash\psi - \frac{g^2}{2}j_\mu^aj^{a\mu}\right).
\eea

As far as the operator content is concerned the building block is the
fundamental Fermi field.  The main condition holds:  in the adjoint
representation of $SU(n)_c$ the color current is the only flavor-singlet,
dimension=1 operator.  In the general OPE operators with canonical dimension=2
can have nonvanishing coefficients, among these there are bilinear and
quadrilinear expressions of the $\psi$-field.  Now we list these.

There are two bilinear, dimension=2 operators, they are the derivative of
the color current - which we expect to have - and an antisymmetric derivative
of the basic fields (which we want to get rid of):
\bean
\dn\jma(x) &\equiv& \dn\bar\psi\lambda^a\gamma_\nu\psi +
                       \bar\psi\lambda^a\gamma_\mu\dn\psi \\
k_{\mu\nu}^a(x) &\equiv& i(\dm\bar\psi\lambda^a\gamma_\nu\psi -
                           \bar\psi\lambda^a\gamma_\nu\dm\psi).
\eean

There are several quadrilinear operators. The chiral symmetry allows
the following $SU(n)_c$-vector and $SU(N)_f$-scalar operators. (The ones
containing the unit in Dirac-space or $\gamma_5$ are either ruled out
by chiral symmetry or can be Fierz-transformed into the following forms)
\bean
l_{\mu\nu}^a(x) &\equiv& (\bar\psi\gamma_\mu\psi)
                         (\bar\psi\lambda^a\gamma_\nu\psi)(x) \\
N_{\mu\nu}^a(x) &\equiv& i\fabc(\bar\psi\gamma_\mu\lambda^b\psi)
                         (\bar\psi\gamma_\nu\lambda^c\psi)(x) \\
n_{\mu\nu}^a(x) &\equiv& \dabc(\bar\psi\gamma_\mu\lambda^b\psi)
                         (\bar\psi\gamma_\nu\lambda^c\psi)(x) \\
\bar{l}_{\mu\nu}^a(x) &\equiv& (\bar\psi\gamma_\mu\Lambda^A\psi)
                         (\bar\psi\lambda^a\gamma_\nu\Lambda^A\psi)(x) \\
M_{\mu\nu}^a(x) &\equiv& i\fabc(\bar\psi\gamma_\mu\Lambda^A\lambda^b\psi)
                         (\bar\psi\gamma_\nu\Lambda^A\lambda^c\psi)(x) \\
\bar{n}_{\mu\nu}^a(x) &\equiv&
\dabc(\bar\psi\gamma_\mu\Lambda^A\lambda^b\psi)
                         (\bar\psi\gamma_\nu\Lambda^A\lambda^c\psi)(x).
\eean

We define the charge conjugation operator as in \myref{chargec}.
Straightforward
calculation shows that among the operators above $k_{\mu\nu}^a$, $l_{\mu\nu}^a$
and $\bar{l}_{\mu\nu}^a$ transform with $-\beta^{ab}$ i.e. they have different
C-parity from the current's and thus do not appear in the OPE.  The operators
$n_{\mu\nu}^a$ and $\bar{n}_{\mu\nu}^a$ are symmetric Lorentz-tensors so
following the general analysis of sec.(\ref{gensec}) only four operators
remain:
\bea
\jma, \dm\jna, N_{\mu\nu}^a, M_{\mu\nu}^a.
\eea


\secII{Perturbative calculation of the OPE}

In the following we will calculate the OPE coefficients up to the first
nontrivial order in the coupling. Of course, a finite order perturbative
approximation is not too conclusive from the nonlocal charge's point of view
but later we will show that even this lowest-order
approximation yields enough information to determine the exact OPE.  The
calculation is standard, the ultraviolet divergences are regulated by
continuing from 2 to $d$ dimensions:
\bea
{\cal L} = \bPsi i\dslash\psi -
\frac{g_0^2(\mu^2)^{1-\frac{d}{2}}}{2}(\bPsi\la\gm\psi)^2.
\eea

The Feynman rules involve only one vertex with four legs. The two incoming
and
two outgoing fermion lines are contracted by the operator:
\bea
-ig^2(\gm \otimes\gum)(\la\otimes\la),
\nn
\eea
and in the many-flavor case
the unit operator contracts the flavor indices.  The
Fourier-transform is defined as:
\bea
\psi(p) \equiv \int\dd^dxe^{ipx}\psi(x).
\nn
\eea

The $\beta$-function of the theory is well known, and can be found in
the literature, e.g. \cite{FNN}:
\bea
\beta(g)\equiv\mu\frac{dg}{d\mu} =
\left(\frac{d}{2}-1\right)g-\frac{q}{2\pi}g^3 +
\frac{q^2}{2\pi^2n}g^5+\ldots
\eea

We computed the 4-point function of the operator product
$\fabc\jmb(x)\jnc(0)$
as $x\rightarrow 0$ up to order $g^2$. The terms which are finite as
$x\rightarrow 0$ yield the four-point function of the renormalized operator,
$N_{\mu\nu}^a(0)$, the infinite terms can be identified with divergent
coefficients times the four-point functions of the other operators:
$\jma(0)$,
$\dm\jna(0)$, $N_{\mu\nu}^a(0)$ and $M_{\mu\nu}^a(0)$. The summary of the
calculation is the following. The amputated two-point function of $\jma(x)$
is:
\bea
\expect{\psi(p_1)\jma(x)\bar{\psi}(p_2)} =
\gamma_\mu\lambda^ae^{i(p_1-p_2)x}(1-g^2\frac{qN}{n\pi}+\ordo{g^4}).
\label{cur2pt}
\eea
We need the two-point function of the OPE at tree level:
\bea
\expect{\psi(p_1)\fabc\jmb(x)\jnc(0)\bPsi(p_2)} =
\nn
\eea
\bea
=\frac{q}{i\pi}\gr\la
\frac{\gmnd x^\rho - x_\mu\delta_\nu^\rho-x_\nu\delta_\mu^\rho}{x^2}
(1+\frac{i}{2}(p_1-p_2)x) + \ordo{|x|}.
\nn
\eea
In view of \myref{cur2pt} this can be written as
\bea
\expect{\psi(p_1)\fabc\jmb(x)\jnc(0)\bPsi(p_2)} =
\eea
\bea
= \frac{q}{i\pi}
\frac{g_{\mu\nu} x^\rho-(\delta_\mu^\rho x_\nu+\delta_\nu^\rho x_\mu)}{x^2}
\expect{\psi(p_1)(\jra(0)+\frac{x^\sigma}{2}\ds\jra(0))\bPsi(p_2)}
+\ordo{|x|}.
\nn
\eea


\begin{figure}[ht]\caption{}

\label{fig:feyn}
\begin{center}
\unitlength=1.00mm
\special{em:linewidth 0.4pt}
\linethickness{0.4pt}
\begin{picture}(100.00,21.00)
\put(15.00,15.00){\circle{2.00}}
\put(14.00,16.00){\line(-1,1){4.00}}
\put(14.00,14.00){\line(-1,-1){4.00}}
\put(20.00,10.00){\circle*{2.00}}
\put(20.00,10.00){\line(-1,1){4.00}}
\put(16.00,16.00){\line(1,1){4.00}}
\put(20.00,10.00){\line(1,0){7.00}}
\put(28.00,10.00){\circle*{2.00}}
\put(28.00,10.00){\line(1,0){6.00}}
\put(50.00,15.00){\circle{2.00}}
\put(60.00,15.00){\circle*{2.00}}
\put(65.00,15.00){\circle*{2.00}}
\put(85.00,15.00){\circle{2.00}}
\put(92.00,10.00){\circle*{2.00}}
\put(92.00,20.00){\circle*{2.00}}
\put(49.00,16.00){\line(-1,1){4.00}}
\put(49.00,14.00){\line(-1,-1){4.00}}
\put(65.00,10.00){\line(0,1){10.00}}
\put(80.00,20.00){\line(1,-1){4.00}}
\put(84.00,14.00){\line(-1,-1){4.00}}
\put(86.00,16.00){\line(3,2){6.00}}
\put(92.00,20.00){\line(1,0){8.00}}
\put(100.00,10.00){\line(-1,0){8.00}}
\put(92.00,10.00){\line(-3,2){6.00}}
\put(55.50,15.00){\oval(9.00,10.00)[]}
\end{picture}
\end{center}
\end{figure}

We now turn to the four-point function of the operator product.  The
tree-level
matrix element is finite as $x\rightarrow 0$, it gives the corresponding
correlation function of $-iN_{\mu\nu}^a(0)$.  At the next order there are
three
different types of diagrams, these are shown in fig.\ref{fig:feyn}.  The
filled
circles stand for the operator of the current and the empty circles denote
the
interaction vertex.  As $x\rightarrow 0$ the graphs of the first type
include the two-point function of the current-current OPE as a subgraph
so we can substitute that result.
It can be shown that the second graph is UV finite,
and thus it yields the corresponding matrix element of
$-iN_{\mu\nu}^a(0)$.  The graphs belonging to the third type are UV
divergent
and give rise to logarithmic $x$-dependence.  Since we expect that the
operator
$-iN_{\mu\nu}^a(0)$ will appear with leading coefficient $1$ we calculated the
difference of the matrix element of the operator product and
$N_{\mu\nu}^a(0)$ (when dealing with diagrams of the third type).

A straightforward application of the Feynman rules shows that the following
one-loop integral need to be calculated:
\bea
f^{\rho\kappa}(p,q,x) \equiv
\idk\frac{ik^\rho}{k^2}\frac{i(p+k)^\kappa}{(p+k)^2}
\left(e^{i(k+q)x}-1\right).
\eea

It is easy to see that $p$ and $q$ do not affect the short-distance
behavior of the integral, the derivatives with respect to these momenta
are UV-convergent and vanishing as $x\rightarrow 0$. Therefore it is
sufficient to consider the integral at arguments $p=q=0$. We obtain:
\bea
f^{\rho\kappa}(0,0,x) =
\frac{i}{8\pi}\left(\frac{2g^{\rho\kappa}}{d-2}-
g^{\rho\kappa}\log(-\pi x^2)-\gE g^{\rho\kappa}-
2\frac{x^\rho x^\kappa}{x^2}\right).
\eea

Here $-\gamma_E$ is the derivative of the $\Gamma$-function at 1.
The UV-divergent term is $x$-independent and is exactly equal to the
divergent
part of the unrenormalized Green's function of $N_{\mu\nu}^a(0)$. The
$x$-independent UV divergences thus cancel when we replace $N_{\mu\nu}^a(0)$
with the renormalized operator. When putting together all the results we
note
that all the possible operators have nonzero four-point function at this
order,
so we can write our result as an operator equation :
\bea
\fabc\jmb(x)\jnc(0) &=& \frac{q}{i\pi}
\frac{g_{\mu\nu} x^\rho-\delta_\mu^\rho x_\nu-\delta_\nu^\rho x_\mu}{x^2}
\left(\jra(0)+\frac{x^\sigma}{2}\ds\jra(0)\right) + \\
\label{pertOPE}
&&+\frac{1}{i}\left(1+\frac{g^2q}{2\pi}(\half+\gE+\log(-\pi\mu^2x^2))\right)
N_{\mu\nu}^a(0)+ \\
&&+\ordo{|x|^{1-0}}. \nn
\eea

There are two important facts to notice in this result.  The only operator
$M_{\mu\nu}^a(0)$ which could have also been present (in the many-flavor CGN)
did not turn up at this order, now the question is whether it can be present
with higher-order coefficient.  In the next section we will show that this can
not be the case:  from the fact that its coefficient is zero at $g^2$-order
follows that it vanishes.  The other issue is the presence of the operator
$N_{\mu\nu}^a(0)$.  Notice that its coefficient is $x$-dependent, i.e. it is
not constant.  However the general analysis of the OPE showed that any
antisymmetric tensoroperator which is independent of the current must have
constant coefficient.  From this it follows, that $N_{\mu\nu}^a(0)$ is {\em not
independent} of the current and its derivative, we can erase it from the list
of the possible operators.  The precise statement is that the classical
curl-free equation survives the quantization in the sense that the two
operators remain linearly dependent.  We give a functional integral argument
for this in the appendix.

\secII{Renormalization group for OPE coefficients}

We now turn to the question of the remaining operator, $M_{\mu\nu}^a(0)$.
The OPE looks like:
\bea
\fabc\jmb(x)\jnc(0) = \CC(x)\jra(0) +
           D_{S,\mu\nu}^{\sigma\rho}(x)S_{\sigma\rho}^a(0)+
           D_{A,\mu\nu}^{\sigma\rho}(x)A_{\sigma\rho}^a(0)+
                 E_{\mu\nu}^{\sigma\rho}M_{\sigma\rho}^a(0),
\eea

where $S$ and $A$ mean the symmetrized and antisymmetrized derivative of the
current (w.r.t. its Lorentz indices).  We found in the general section that the
coefficient $E$ must be constant.  Since the currents are not renormalized, the most general
renormalization which involves the four operators is:
\bea
\left(
\begin{array}{c}
\jma \\ S_{\mu\nu}^a \\ A_{\mu\nu}^a \\ M_{\mu\nu}^a
\end{array} \right)_{Ren.} =
\left(\begin{array}{cccc}
1 & 0 & 0 & 0 \\
0 & 1 & 0 & 0 \\
0 & 0 & 1 & 0 \\
0 & 0 & Z_{MA} & Z_{MM}
\end{array}\right)
\left(\begin{array}{c}
\jma \\ S_{\mu\nu}^a \\ A_{\mu\nu}^a \\ M_{\mu\nu}^a
\end{array} \right)_0\,.
\eea

When calculating the OPE coefficients perturbatively one obtains an
explicit
dependence of the Green's functions on the renormalization scale, $\mu$. The
renormalization group equations ensure that the physical quantities do not
depend on this scale if one also considers the implicit $\mu$-dependence of
the couplings. By forming Green's functions from both sides of the OPE one
can write RG equations for the OPE coefficients. Straightforward application
of the operator ${\cal D}\equiv\mu\frac{d}{d\mu}$
to both sides yields:
\bea
0 &=& {\cal D}\CC(x) \\
0 &=& {\cal D}D_{S,\mu\nu}^{\sigma\rho}(x) \\
0 &=& {\cal D}D_{A,\mu\nu}^{\sigma\rho}(x) +
       E_{\mu\nu}^{\sigma\rho} {\cal D}Z_{MA} -
       E_{\mu\nu}^{\sigma\rho} \frac{Z_{MA}}{Z_{MM}}{\cal D}Z_{MM} \\
0 &=& {\cal D}E_{\mu\nu}^{\sigma\rho} +
      E_{\mu\nu}^{\sigma\rho}{\cal D}\log(Z_{MM}).
\eea
The tensor structure of $E_{\mu\nu}^{\sigma\rho}$ is
$(\delta_\mu^\sigma\delta_\nu^\rho - \delta_\nu^\sigma\delta_\mu^\rho)E$,
from which the last equation yields:
\bea
{\cal D}\log(EZ_{MM}) = 0.
\label{EZMM}
\eea

Suppose that we know the full perturbative series of $E$, and the anomalous
dimension of $M_{\mu\nu}^a$ is also given as a power series in the coupling:
\bea
{\cal D}\log(Z_{MM}) \equiv \gamma_{MM} &=&
\eta_1 g^2 + \eta_2 g^4 + \ldots \\
E &=& E_0g^{2\alpha}(1+E_1g^2+E_2G^4+ \ldots).
\eea

$E_i$ are independent of the space-time difference therefore they can not
depend on $\mu$, either, since there is no other dimensionful quantity
and the RG-derivative becomes very simple:
\bea
{\cal D}\log(E Z_{MM}) = 2g\beta(g)\left(\frac{\alpha}{g^2}+E_1+\ldots
\right)+\gamma_{MM}.
\eea

The power expansion of the RG equation \myref{EZMM} then gives:
\bea
0 &=& 2\beta_0\alpha+\eta_1+\ordo{g^2} \\
\alpha &=& -\frac{\eta_1}{2\beta_0}.
\eea

The important consequence of this result is that the exponent of the
leading perturbative
behavior of the coefficient $E$ is known if we calculate the 1-loop $\beta$-
and $\gamma$-functions.  Having calculated this number, $\alpha$ we obtain the
following information.  If $\alpha$ is {\em not} a nonnegative integer then the
only possible solution for the RGE is $E=0$ since one can not obtain
this kind of
behavior in perturbation theory.  If $\alpha$ is a nonnegative
integer then it shows the order of
perturbation theory where one first obtains a
nonzero result when calculating the OPE
coefficient.  This is most informative in the case when one calculates the
coefficient up to $\alpha$th order and obtains zero; then conclusion is then
that it remains zero to all orders.

It is a straightforward exercise to calculate $\gamma_{MM}$ at first order, we
get:  $\eta_1 = \frac{q}{\pi}$.  From this we obtain $\alpha=1$.  In the
previous section we found that $E=0$ at the order of $g^2$ which shows
that the OPE coefficient remains zero to any order. Thus the operator
$M_{\mu\nu}^a(0)$ {\em can not enter the OPE}.

\secII{Summary}

The perturbative calculation in the one-flavor and many-flavor CGN model showed
that in our OPE only the color current and its derivative is present.  With the
help of \myref{pertOPE} and \myref{qucurlfree} we can determine the scalar OPE
coefficients:
\bea
\gamma_2&=& -\gamma_1 = \lambda = -\frac{q}{i\pi} \\
\delta_A &=&i\left(\frac{2}{g^2}
- \frac{q}{\pi}(\log(-\pi\mu^2x^2)+\gamma_E+\half)\right).
\eea

This can be transformed into a more transparent form if we use the
asymptotic
freedom and introduce the ($x$-dependent) running coupling:
\bea
\frac{1}{\bar{g}^2} = -\frac{q}{2\pi}\log(-\Lambda^2x^2)-
\frac{q}{\pi n}\log(-\log(-\Lambda^2x^2))+
\frac{q}{2\pi}\log(-\mu^2x_0^2)+const.
\nn
\eea

Here $\Lambda$ is the dimensional transmutation parameter. Then the OPE
coefficient (in $\overline{MS}$ scheme) is:
\bea
\delta_A = \frac{q}{i\pi}\log(-\Lambda^2x^2)+
\frac{2q}{i\pi n} \log(-\log(-\Lambda^2x^2))+\frac{q}{\pi i}(\half-\log(4)).
\eea

In the one-flavor CGN-model the existence of the non-local charge is
essentially ensured by the kinematics (similarly to the $O(n) \sigma$-models):
simply there is no extra operator which could appear in the OPE. This is a
well-known result \cite{AbAb}. The novelty in our treatment of the MCGN is that
in this case the general OPE does not yield enough information and we had to
use a dynamical argument to rule out the extra operator which could have been
present a priori.

\secI{Conclusion}

In the present paper we discussed two subjects. By reviewing previous results
obtained in integrable models we determined a general sufficient criterion for
the existence of conserved non-local charges and Yangian algebra in a class
of 2d quantum field theories. We showed the crucial role of the current-current OPE
and
pointed out how the most general form (allowed by general principles) of the
OPE-coefficients ensures the existence and conservation of the
non-local charge provided
there are no "extra" operators in the Wilson expansion. It turned out that the
properties and the coefficients of these operators are also highly constrained
which may help in dealing with them in concrete theories.

We employed these
general results to compare the one- and multiflavor CGN model. As it was found
previously \cite{AbAb} the one-flavor model belongs to the trivial case of the
analysis: there are too few operators and the algebra of the currents must
close on themselves. This is not the case in the multiflavor model where the
flavor-space increases the number of degrees of freedom and the new operators
could in principle invalidate the conservation of the nonlocal charge. We had
to take a closer look at the OPE for which we used renormalized perturbation
theory improved by renormalization group. The vanishing of the extra
operator at one loop level was promoted by the RG-arguments to all orders thus
answering the crucial question: the OPE of the current closes on themselves
in the multiflavor model too.

Our results show that although the particle spectrum of the one- and the
multiflavor chiral model is different \cite{PolyWieg} nevertheless the
same Yangian symmetry algebra is realized. This fact points to the advantage
of the operator-algebra approach: one can study the symmetry structure without
having to refer to the spectrum of the theory.

The proof of the existence of the Hopf-algebra symmetry generated by non-local
charges is a first step in proving the integrability of a model. We think
that the consequences on the mass spectrum, S-matrix and correlation functions
can be analyzed similarly to the models studied previously, e.g
\cite{Luscher,Abdalla,BerLecl}.

I wish to thank J. Balog for the introduction into this field and his
continuous support and gratefully acknowledge discussions with P. Forg\'acs.

\secI{Appendix: The curl-free equation}

As it is discussed in \cite{PolyWieg,Destri} the current-current
interaction of the multiflavor CGN model can be replaced by an equivalent
vector-boson mediated one:
\bea
S = \int d^2x \left(
\bPsi i\dslash\psi - A_\mu^aj^{\mu a} +\frac{1}{2g^2}A_\mu^a A^{\mu a}
\right)
\eea
The functional integral (which should be defined in Euclidean space by Wick
rotation) over the auxiliary field $A_\mu^a$ is purely Gaussian thus the quantum
equivalence clearly holds. On the other hand one can perform the fermionic
integral first; leading to the effective action of the gauge fields
(which may be viewed as a classical background). The breakdown of the axial
invariance of the effective action yields the axial anomaly equation:
\bea
D_\mu j_5^\mu = \frac{Nq}{2\pi n}\epsilon^{\mu\nu}F_{\mu\nu}
\eea
Here $j_5^{\mu a}=-\epsilon^{\mu\nu}j^a_\nu$ is the axial current in 2d and
$F_{\mu\nu}$ is the field-strength tensor corresponding to $A_\mu^a$. This
equation is obtained in any gauge-invariant regularization. By
considering this equation inside a correlation function involving fermionic
fields one can perform the integral over the vector field first and
an operator equation in the original purely fermionic model is obtained.
Note that the integral does not cause any problem in the quadratic
part of $F_{\mu\nu}$ because of the antisymmetry of the color
structure constant,
$\fabc$. Evaluating the functional integral over the vector field
then yields:
\be
\label{qucurlfree}
\left(1+g^2\frac{qN}{\pi n}\right)(\dm\jna-\dn\jma)+
2g^2\left(1+g^2\frac{qN}{2\pi n}\right)i\fabc\jmb\jnc = 0.
\label{qcurl1}
\ee
Note that our vector-boson theory does not require a gauge-invariant
scheme and by a suitable
finite renormalization we can change the anomaly equation such that the
classical curl-free equation is not modified. One can therefore {\em define}
the quantum CGN-model keeping
\be
\dm\jna-\dn\jma+2g^2i\fabc\jmb\jnc = 0.
\label{qcurl2}
\ee
Whether \myref{qcurl1} or \myref{qcurl2} is chosen as the
renormalization scheme the lesson is the same: the operator $N_{\mu\nu}^a$ is
not independent of the antisymmetrized derivative of the current.

\end{document}